\documentclass[sn-mathphys]{sn-jnl}
\usepackage{url,hyperref,lineno,microtype}
\usepackage[onehalfspacing]{setspace}
\usepackage{multirow}
\usepackage{xspace}

\jyear{2021}
\theoremstyle{thmstyleone}

\theoremstyle{thmstyletwo}

\theoremstyle{thmstylethree}

\newcommand{\hyppo}{\textsc{Hyppo}\xspace}
\newcommand{\pp}{\mathbf{p}}
\newcommand{\ww}{\mathbf{w}}
\newcommand{\Dm}{\mathcal{D}}

\newcommand{\noisedim}{\texttt{noise\_dim}}

\newcommand{\lrgen}{\texttt{lr\_gen}}
\newcommand{\lrdisc}{\texttt{lr\_disc}}
\newcommand{\lsgen}{\texttt{gen\_layersize}}
\newcommand{\lsdisc}{\texttt{disc\_layersize}}

\newcommand{\HW}{\textsf{Herwig}\xspace}
\newcommand{\pt}{\ensuremath{p_\text{T}}\xspace}
\newcommand{\zmumu}{\ensuremath{Z/\gamma*\to\mu\mu}\xspace}

\begin{document}

\title[Optimizing GANs for HEP Simulations]{Hyperparameter Optimization of Generative Adversarial Network Models for High-Energy Physics Simulations} 

\author*[1]{\fnm{Vincent} \sur{Dumont}}
\email{vincentdumont11@gmail.com}

\author[1]{\fnm{Xiangyang} \sur{Ju}}
\equalcont{These authors contributed equally to this work.}

\author[1]{\fnm{Juliane} \sur{Mueller}}
\equalcont{These authors contributed equally to this work.}

\affil*[1]{\orgname{Lawrence Berkeley National Laboratory}, \orgaddress{\city{Berkeley}, \postcode{94720}, \state{CA}, \country{USA}}}

\abstract{The Generative Adversarial Network (GAN) is a powerful and flexible tool that can generate high-fidelity synthesized data by learning. It has seen many applications in simulating events in High Energy Physics (HEP), including simulating detector responses and physics events. However, training GANs is notoriously hard and optimizing their hyperparameters even more so. It normally requires many trial-and-error training attempts to force a stable training and reach a reasonable fidelity. Significant tuning work has to be done to achieve the accuracy required by  physics analyses. This work uses the physics-agnostic and high-performance-computer-friendly hyperparameter optimization tool \hyppo to optimize and examine the sensitivities of the hyperparameters of a GAN for two independent HEP datasets. This work provides the first insights into efficiently tuning GANs for Large Hadron Collider data. We show that given proper hyperparameter tuning, we can find GANs that provide high-quality approximations of the  desired quantities. We also provide guidelines for how to go about GAN architecture tuning using the analysis tools in \hyppo.}


\maketitle

\section{Introduction}

Event generation and detector simulation are essential for physics analyses at the Large Hadron Collider, but are also computationally expensive. Different Machine Learning-based generative models are exploited to reduce the computational cost. Those generative models can be classified into four categories: (1) Variational Autoencoders~\citep{1312.6114K}, which learn a stochastic map from the data space to a latent space and back, preserving the statistics of the latent space and data space; (2) Normalizing Flows~\citep{pmlr-v37-rezende15} use invertible transformations so that the probability density can be computed and the generator is optimized using the log likelihood; (3) score-based generative models~\citep{ho.scorefunction,song.scorefunction}, which generate samples from noise by repeatedly perturbing the data with a diffusion equation, and learning to reverse the perturbation via estimating the diffusion function; (4) Generative Adversarial Networks (GAN)~\citep{NIPS2014_5ca3e9b1}, which optimize the generator network by means of an auxiliary network (‘discriminator’) that tries to classify generated examples from real examples. GAN will remain as an important generative model in High Energy Physics because of its unique features.

GANs have been used in many aspects of High Energy Physics to accelerate computationally intensive physics simulations. Different experiments at the Large Hadron Collider (LHC) are using GANs to enable fast simulations for their calorimeters~\citep{Paganini:2017dwg, Chekalina:2018hxi,ATL-SOFT-PUB-2018-001,Deja:2019vcv,Erdmann:2018jxd}. In addition, GANs are used to simulate physics events for data analyses~\citep{Otten:2019hhl,Butter:2019cae,DiSipio:2019imz,Choi:2021sku,Butter:2022rso}, unfolding detector-level measurements to parton-level observables~\citep{Datta:2018mwd}, and compressing Parton distribution function (PDF) sets~\citep{Carrazza:2021hny}. However, these GANs have not yet reached  production level fidelity and reliability, and much work remains to be done to efficiently train GANs and tune their architectures.

It is very hard to train a GAN and even harder to achieve a desired performance. In the field of image processing, lots of research is targeted at improving the stability of training GANs by modifying the training algorithm~\citep{2015arXiv150605751D,timGanTips}. However, these techniques may not be applicable to sparse and  high-dimensional scientific data. \citep{Oliveira:DLPS2017} offer some best practices for training a GAN for scientific data, but they do not perform a thorough analysis of the impact of the GAN's architecture on the model performance. The architecture  defined by hyperparameters such as layer sizes determines how well the model can perform. For example, a highly complex architectures trained on a small dataset may quickly lead to overfitting of the data, whereas a too simplistic architecture may lead to bad predictive performance.

Different methods for tuning deep learning (DL) model architectures have been introduced in the literature. Trial and error and random sampling methods have been widely used~\citep{Bergstra2012}. However, these methods are  not efficient and often lead to suboptimal solutions. More advanced tuning methods exist. These include MENNDL~\citep{MENNDL}, which uses an evolutionary optimization approach; DeepHyper~\citep{DeepHyper}, which is based on Bayesian optimization methods; Hyperband~\citep{Hyperband}, which aims at accelerating random search by implementing early stopping criteria; as well as  Optuna~\citep{optuna_2019} and Hyperopt~\citep{Bergstra2013}, which both offer random sampling and  Tree-structured Parzen Estimator (TPE) methods. In addition, \citep{HPOreview} reviews on hyperparameter optimization (HPO) methods and \citep{2022arXiv220106433S} compares these methods with open benchmarks.  The biggest drawback of these methods is that they do not take into account the prediction variability of the  models during the hyperparameter search. This can lead to identifying architectures that do not offer reliable performance and produce predictions with large variabilities. 

\hyppo~\citep{hyppo} is a software for tuning DL model architectures (hyperparameters) by building surrogate models that consider both the prediction accuracy and prediction variability, thus leading to DL models that make reliable predictions. In addition, \hyppo is High-Performance-Computing (HPC) friendly. It uses the Message Passing Interface to communicate between computing nodes and allows users to perform large scale HPOs in a reasonable time. Like  DeepHyper, Optuna and Hyperopt, \hyppo relies on surrogate models (for details, see Section~\ref{ssec:hyppo}). Surrogate models have been widely used in the literature for solving computationally expensive black-box optimization problems (problems for which an analytic description of the objective function does not exist), see~\citep{Booker,Audet2017,surrogate4opt}. 

In this work, we leverage \hyppo to perform HPO of GANs that are trained on two distinct yet representative   simulated 
LHC datasets with the goal to  achieve highly accurate models and we analyze the models' performance sensitivities  with regard to the hyperparameters. We show that for both datasets, we can achieve highly accurate results with our tuned GANs.  

The remainder of this article is organized as follows. In Section~\ref{sec:physicsGANstuff}, we  describe how we train and tune the GANs used in our study and we define the performance metrics used to assess the model performance. Section~\ref{sec:hyppo} provides a brief description of the hyperparameter optimization task and the \hyppo optimizer.

Section~\ref{sec:exp} discusses the numerical experiments for our two case studies as well as  the HPO  results and outcomes of the sensitivity analysis. Conclusions and future research opportunities are provided in Section~\ref{sec:conclude}.

\section{Training and Optimizing GANs for LHC Event Generation Applications}
\label{sec:physicsGANstuff}
In this section, we provide details about the GANs and their hyperparameters used in our case studies. We also explain  the performance  metrics we use to compare different architectures. 

\begin{figure}[ht]
\begin{center}
\includegraphics[width=0.49\columnwidth]{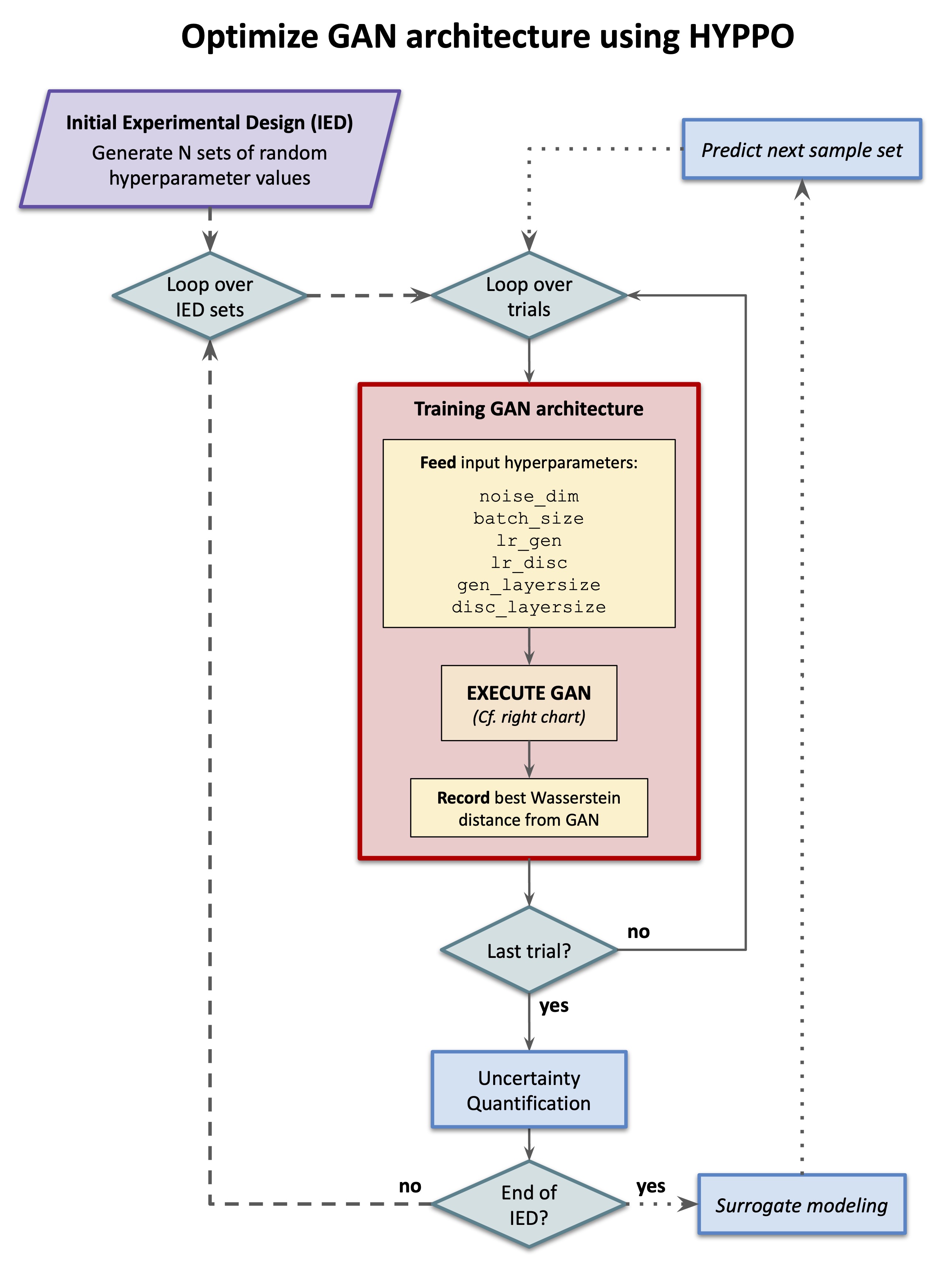}
\includegraphics[width=0.49\columnwidth]{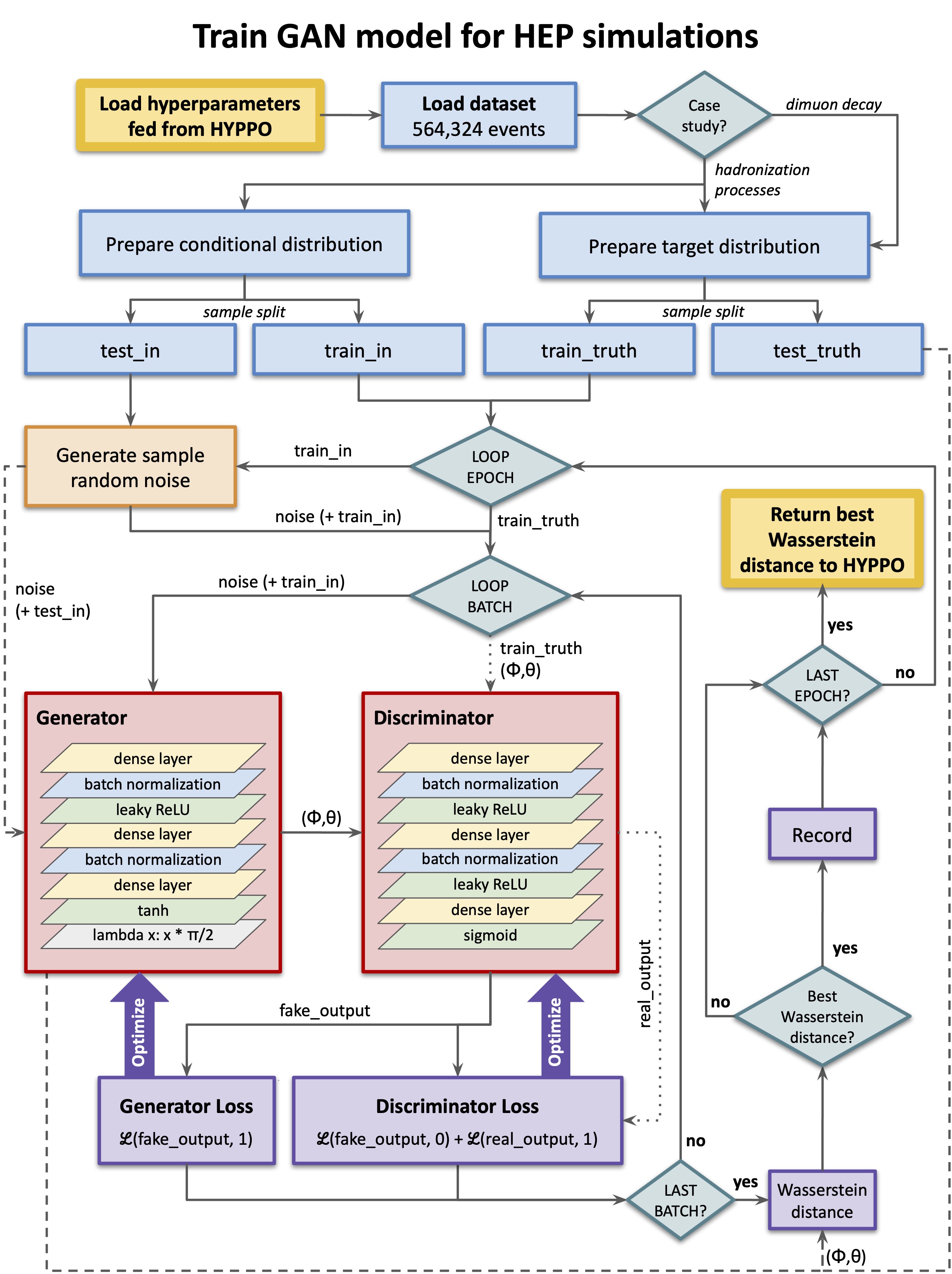}
\end{center}
\caption{Flowchart representing the general hyperparameter tuning (left) and training (right) workflow.  }
\label{fig:flowchart}
\end{figure}

In Fig.\,\ref{fig:flowchart}, we present the workflows used in our study for training the GAN (right) and optimizing its architecture (left).
 The left image  shows a rough feedback loop in which the GAN architecture is automatically adjusted using \hyppo's optimization approach (for details, see Section~\ref{ssec:hyppo}). The right figure shows the details of the GAN training that takes place in each iteration of \hyppo when a hyperparameter set is evaluated for its performance. The right image is a detailed description of  the red box in the left image. 

The GAN  (right figure) is composed of a Generator Network (the Generator) and a Discriminator Network (the Discriminator). For the first case study (see Section~\ref{sec:herwig_data}),  the Generator takes as inputs the incoming particle's four-vector and  random values drawn from a Gaussian distribution with a mean of zero and a standard deviation of one (i.e., the Gaussian noise), and it outputs two physics variables (spherical coordinates of one outgoing particle; the second particle's data can be inferred using energy conservation laws). In this way, the output variables are \textit{conditioned} on the incoming particle's four-vector.
In the second case study (see Section~\ref{dimuon}), no input four-vector information is fed as initial training data and only random noise will be used. For this case, the output variables correspond to the kinematics of the two muons.

The dimension of the random values is called \textit{noise dimension}, \noisedim, which is a tunable hyperparameter. Both the Generator and the Discriminator use multi-layer perceptrons (MLPs) with two layers each. The layer sizes of both the Generator (\lsgen) and the Discriminator (\lsdisc) are tunable hyperparameters. Batch normalization is applied after the dense layers, followed by a \textsc{Leaky ReLU}~\citep{leakyReLU} activation function. For the generator, a \texttt{tanh} activation function is used to ensure that the Generator output data are between -1 and 1, and, finally, a \texttt{lambda} function is implemented as the final layer to scale the Generator outputs back to their true ranges so that they can be directly compared to the ground truth  simulation data. Similar to the Generator, the Discriminator also corresponds to a two-layer MLP with the difference that the final activation function is a \texttt{sigmoid} function that yields a score ranging from zero to one. The higher the score is, the more likely is the Discriminator to  think that the input values are from the true distribution. The Discriminator and the Generator are trained separately and alternately by two separate Adam optimizers~\citep{2014arXiv1412.6980K} that minimize the binary cross-entropy, each with their own learning rate, for  $M$ epochs. The learning rates of both the Generator (\lrgen) and the Discriminator  (\lrdisc) are  tunable hyperparameters. 

Table~\ref{table:hps_range} shows the hyperparameters to be tuned and their search ranges. Hyperparameters can be any natural number in the predefined range from \textit{Lower} to \textit{Upper} and the natural number is scaled by a scaling factor \textit{Scaling}. By doing this, continuous hyperparameter values are automatically discretized and can thus be treated in the same way as the discrete ones.

\begin{sidewaystable}
\sidewaystablefn%
\begin{center}
\begin{minipage}{\textheight}
\caption{Hyperparameters we  optimize in our study. The lower and upper bound for each hyperparameter as well as the scaling factors are shown  for both the broad search (with identical search ranges for both case studies) and the focus search (with the values depending on the results from the broad search).}\label{table:hps_range}
\begin{tabular}{ |c|c|c|c|c|c|c|c|c|c| }
\hline
\multirow{2}{*}{\bf Hyperparameter} & \multicolumn{3}{c|}{\bf Broad Search}  & \multicolumn{3}{c|}{\bf Study 1 - Focus Search} & \multicolumn{3}{c|}{\bf Study 2 - Focus Search}\\
\cline{2-10}
& {\it Lower} & {\it Upper} & {\it Scaling} & {\it Lower} & {\it Upper} & {\it Scaling} & {\it Lower} & {\it Upper} & {\it Scaling} \\
\hline
\texttt{noise\_dim}      & 1 & 100 &    4 &   1 &  10 &    1 &  1 &  20 &    1 \\
\texttt{batch\_size}     & 8 &  64 &   32 & 640 & 960 &    1 & 30 &  55 &   10 \\
\texttt{lr\_gen}         & 1 & 200 & $5\times10^{-5}$ &   1 &  50 & $10^{-7}$ & 10 & 500 & $10^{-6}$ \\
\texttt{lr\_disc}        & 1 & 200 & $5\times10^{-5}$  &  40 &  70 & $10^{-5}$ &  1 & 200 & $10^{-5}$ \\
\texttt{gen\_layersize}  & 8 & 128 &    8 &  60 & 100 &    1 & 50 & 250 &    1 \\
\texttt{disc\_layersize} & 8 & 128 &    8 & 240 & 320 &    1 & 16 & 256 &    4 \\
\hline
\end{tabular}
\end{minipage}
\end{center}
\end{sidewaystable}

We start the hyperparameter optimization with a broad range as listed in Table~\ref{table:hps_range}, namely the \textit{broad search}. And we use only a small number of training epochs (20) to get a rough idea of how the promising hyperparameters distribute. Training models over a small number of epochs allows us to search a large variety of hyperparameters within a relatively short amount of time (i.e. within minutes), and therefore we can evaluate a larger number of hyperparameter sets within a few hours. After the broad search, a narrower search range is defined based on the hyperparameter sets resulting in good performance. Another HPO with more training epochs (up to 1000) within the narrower bounds, namely the \textit{focus search}, is conducted to find the best model. More details about the HPO are provided in Section~\ref{sec:hyppo}.

The performance of the GAN is measured by the ``Wassertstein distance'' (WD)~\cite{Komiske:2019fks,Cai:2021hnn}, which calculates the minimum work needed to transfer one distribution into another. It can be used to measure the difference between two collision events~\citep{Komiske:2019fks} (here, between the simulated and the GAN-generated distributions). The WD enables us to quantify and optimize the GAN performance.

We monitor the Wasserstein distance in each training epoch. At the end of each epoch, we compute the Wasserstein distance for each quantity (output) the Generator predicts and we average these Wasserstein distances over all quantities. After the last epoch is completed, we select the model that achieved  the lowest average Wasserstein distance  and we use this value as our performance metric of the corresponding hyperparameters,  which is then used to inform the HPO's iterative sampling decisions. Note that since we do not minimize the Wasserstein distance during training, the model with the lowest Wasserstein distance is not necessarily found in the last epoch.

\section{Bi-level Optimization for Architecture Tuning}
\label{sec:hyppo}

In this section, we provide a description of the optimization model for tuning the GAN model architecture and  of our hyperparameter optimization method \hyppo, which we use in this study. 

\subsection{GAN Architecture Optimization Model}
The architecture  tuning problem is a bilevel optimization problem:
\begin{align}
 \min_{\pp, \ww} & \quad U(\pp,\ww^*; \Dm_{\text{val}}) \\
    \text{s.t. }& \pp\in\Omega\\
    & \ww^* \in\min_{\ww\in W} \ell(\ww; \pp, \Dm_{\text{train}}),\label{eq:subp}
 \end{align}
i.e., it is an optimization problem with an embedded \textit{lower level} optimization problem. Here, $\pp$ are our hyperparameters of the GAN  and $\Omega$ is the domain over which they are defined (see Table~\ref{table:hps_range}). Most hyperparameters are by definition restricted to integer values (e.g., the number of layers, nodes per layer, batch size). Other hyperparameters such as the learning rates are continuous. However, for continuous parameters we only allow a limited number of possible options because we assume that (1) small changes of these hyperparameters do not change the performance of the model significantly, and (2) in widely-used trial-and-error settings when  hyperparameters are adjusted by hand, only a limited number of options is considered. Thus, we believe that our assumption about limiting the possible values for each hyperparameter to a finite set  is reasonable and practical. 

In order to assess the performance $U$ (Wasserstein distance) of a hyperparameter vector $\pp$, we have to solve the lower level  optimization problem (Eq.~\ref{eq:subp}) in which we train the corresponding model  by optimizing its weights and biases $\ww$ over the training dataset $\Dm_{\text{train}}$ (we minimize the cross-entropy). Only after we have trained the architecture can we determine its Wasserstein distance $U$ by evaluating the trained model on the validation dataset $\Dm_{\text{val}}$.

Solving the lower level problem (Eq.~\ref{eq:subp}) can be computationally very expensive depending on the size of the architecture and the amount of training data used. Thus, the goal is to find a (near-)optimal solution by evaluating  $U$ for as few hyperparameters  as possible.  To this end, we use a hyperparameter optimization method we previously developed~\citep{gw} and which we implemented in the software \hyppo~\citep{hyppo}. This method is based on surrogate models and uses active learning strategies to guide the search for optimal hyperparameters, thus minimizing the number of hyperparameters that must be tried.

\subsection{Hyperparameter Optimization with \hyppo} \label{ssec:hyppo}

Training a DL model (solving the optimization problem in~(\ref{eq:subp}) is usually done with a stochastic optimizer, such as stochastic gradient descent, due to the vast number of parameters $\ww$ that must be optimized, which prohibits the use of exact methods. However,  training the same architecture multiple times yields different outcomes for $\ww^*$ due to the use of stochastic optimizers, and thus we can obtain different Wasserstein distances  for the same hyperparameter set. Not taking into account this variability during HPO can potentially lead to selecting DL model architectures that are unreliable with regard to their  performance. In order to address this issue in \hyppo, we take an ensemble approach  to assess the  variability of the Wasserstein distance and to find DL model architectures with robust performance. To this end, we train each architecture multiple times (we solve the optimization problem in~(\ref{eq:subp}) multiple times) and we compute the sample average of  the Wasserstein distances   which we then use as  objective function value $U$.

In order to minimize the number of architectures $\pp$ we evaluate during the optimization, we use a surrogate model based optimization algorithm (see also the flowchart on the left in  Figure~\ref{fig:flowchart}). This method starts by creating an initial experimental design $\mathcal{P}=\{\pp_0, \ldots, \pp_{N}\}$ with $N$ different hyperparameter sets. We evaluate the corresponding Wasserstein distances $\mathcal{U}=\{U(\pp_i)\}_{i =0}^{N}$ as described above, where the dependence of $U$ on $\ww$ and $\Dm_{\text{val}}$ is implied. Given these input-output pairs,  we build a surrogate model of $U$: $U(\pp) = s(\pp) +e(\pp)$, where $e(\pp)$ denotes the difference between the true function $U$ and the surrogate model $s$. Different types of surrogate models can be used, and we implemented both Gaussian process (GP) models and radial basis function (RBF) models. We then use the surrogate $s$ to select the next point $\pp_{\text{new}}\in\Omega$ and evaluate $U_{\text{new}}=U(\pp_{\text{new}})$. We update the sets $\mathcal{P} = \mathcal{P} \cup \{\pp_{\text{new}}\}$ and $\mathcal{U} = \mathcal{U}\cup \{U_{\text{new}}\}$ and we update the surrogate model with the new data. The process iterates until we have reached a predefined budget of function evaluations. For further details about this  active learning method, we refer to~\cite{gw}.

\section{Numerical Experiments}
\label{sec:exp}

In this section, we provide a brief desctiption of our datasets and a thorough analysis of our findings. For both case studies, we compare our results obtained with \hyppo to those obtained by randomly sampling the hyperparameter space. 

\subsection{Case Study 1: Simulation of Hadronization Processes}
\label{sec:herwig_data}

Hadronization is a complex quantum process whereby quarks and gluons are converted to hadrons. Current hadronization models are physics-inspired phenomenological models with many parameters that must be tuned to experimental measurements.~\cite{Ghosh:2022zdz} simulated the hadronization process with the \HW\ generator and trained a GAN to learn the kinematic properties of the decay products of the hadronization cluster, an intermediate object taking as inputs the kinematic and flavor information from quarks and gluons and producing the kinematics of hadrons. There are some discrepancies between GAN-generated and \HW-simulated events since an extensive hyperparameter optimization was not performed. Here, we use the same dataset and tune the same GAN architecture as used by~\cite{Ghosh:2022zdz} to study the sensitivities and importance of individual GAN hyperparameters and we optimize those hyperparameters for better performance.

\subsubsection{Simulated HEP Events}

The input dataset consists of 564000 events produced using the cluster model from~\cite{Webber:1983if} decaying into two pions. For a relatively fast turnaround, 5000 events are used for validation. The cluster four momentum is the conditional input to the Generator, which then predicts the polar angle, $\theta$, and azimuthal angle, $\phi$, of the leading~\footnote{Pions are sorted by their \pt} pion's momentum in the spherical coordinate system in the cluster frame in which the two pions are created back-to-back. The four momentum of the other pion is inferred through the four momentum conservation.

\subsubsection{Broad Search: Comparing Random Sampling and Surrogate Model Optimization}

The hyperparameter search space for the broad search is defined in Table~\ref{table:hps_range}. We perform  both the random hyperparameter sampling and the surrogate model guided HPO with \hyppo  and we executed the runs on NERSC's Cori Graphics Processing Unit (GPU) nodes. In order to take into account the performance variability associated with a given model architecture, we performed multiple independent trainings (that is, using different initializations for the weights and biases when solving the lower level optimization problem). Hereafter each training is referred to as a ``trial''. We then compute the mean and standard deviation over the Wasserstein distances of all trials for a given hyperparameter set. The standard deviation (denoted as $1\sigma$ deviation) is also referred to as \textit{model variability}. A single GPU processor is used for each trial. 

The random  sampling  was performed using 100 randomly selected hyperparameter sets. The search was repeated 20 independent times, leading to a total of 2000 unique hyperparameter sets evaluated. When using the surrogate model guided search  (i.e., using GPs and RBFs, respectively), an initial experimental design (IED) with  $N=10$ randomly selected hyperparameter sets is generated and evaluated before proceeding with the surrogate modeling for another $\sim$ 90 hyperparameter evaluations. We ran 20 independent surrogate model based HPOs to average out the stochasticity associated with the algorithm. The same IED is used for both surrogate-based HPO methods to ensure both approaches start with identical conditions.

\begin{figure}[ht]
\includegraphics[width=\linewidth]{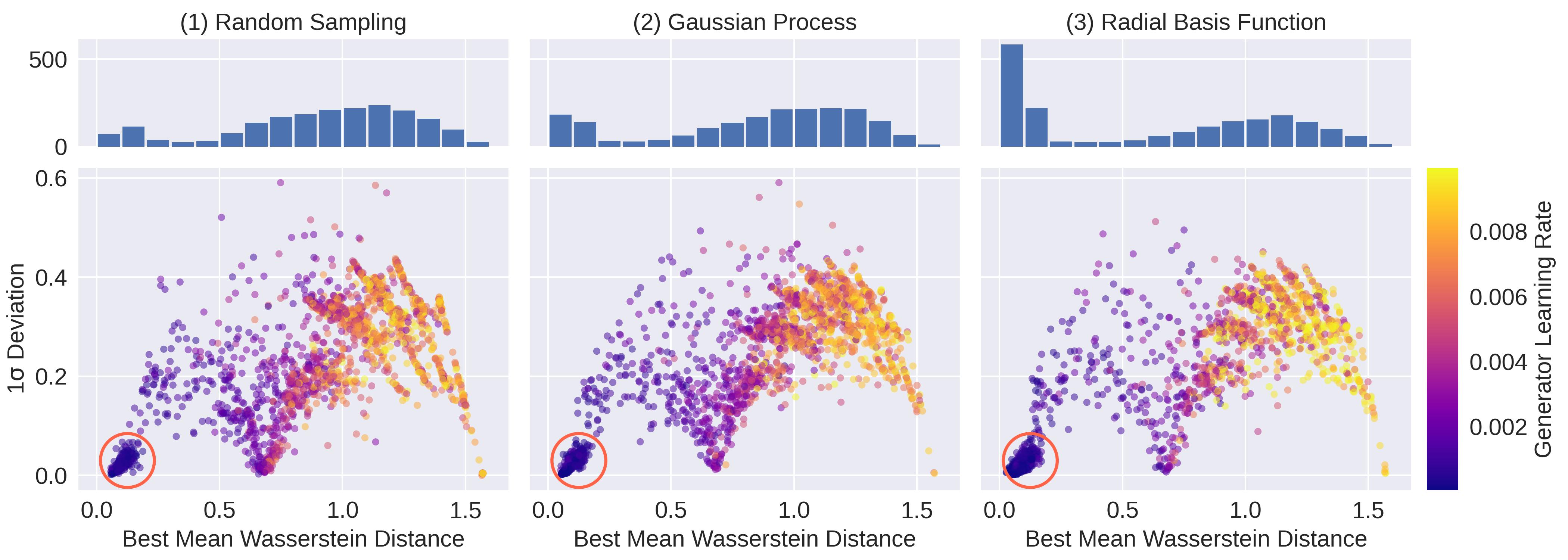}
\caption{Solutions sampled by the three different HPO approaches: (1) random hyperparameter sampling, (2) Gaussian Process surrogate modeling, and (3) Radial Basis Function surrogate modeling.  For each approach, a total of 20 independent HPO runs was performed with over 100 different GAN architectures evaluated in each run. The scatter plots show all architectures (about 2000 for each method) plotted according to their best  averaged Wasserstein distance and corresponding  variability (${1\sigma}$ deviation). A cluster of promising solutions is indicated by the red circle for illustrations. The histograms above the scatterplots represent the number of hyperparameter sets sampled across Wasserstein distances.}
\label{fig:case1_solutions}
\end{figure}

Fig.\,\ref{fig:case1_solutions} shows  the solutions sampled during all 20 optimization repetitions  carried out for each of the three different HPO approaches (random sampling, GP surrogate modeling, and RBF surrogate modeling) using five training  trials for each architecture.  
For each sampled hyperparameter set, we show the averaged Wasserstein distance and the corresponding  variability ($1\sigma$ deviation) computed over all trials  as scatterplots (each point represents an architecture). The goal is to find architectures  with low Wasserstein distance and low variability. The different colors represent the generator learning rate. We can see that lower generator learning rates correspond to models with better Wasserstein distances. The big red circles indicate the best-performing architectures and we can see that all three sampling methods find solutions in the desirable area because each method has a cluster of solutions inside of the circle.   

The histograms above the scatter plots show the number of architectures sampled across Wasserstein distances. We can clearly see that the RBF-based HPO method samples significantly  more architectures in the good Wasserstein distance regime than both the  GP-based and random sampling. This indicates that using the RBF-based optimization is effective  for this case. 

\begin{figure}[ht]
\begin{center}
\includegraphics[width=0.8\columnwidth]{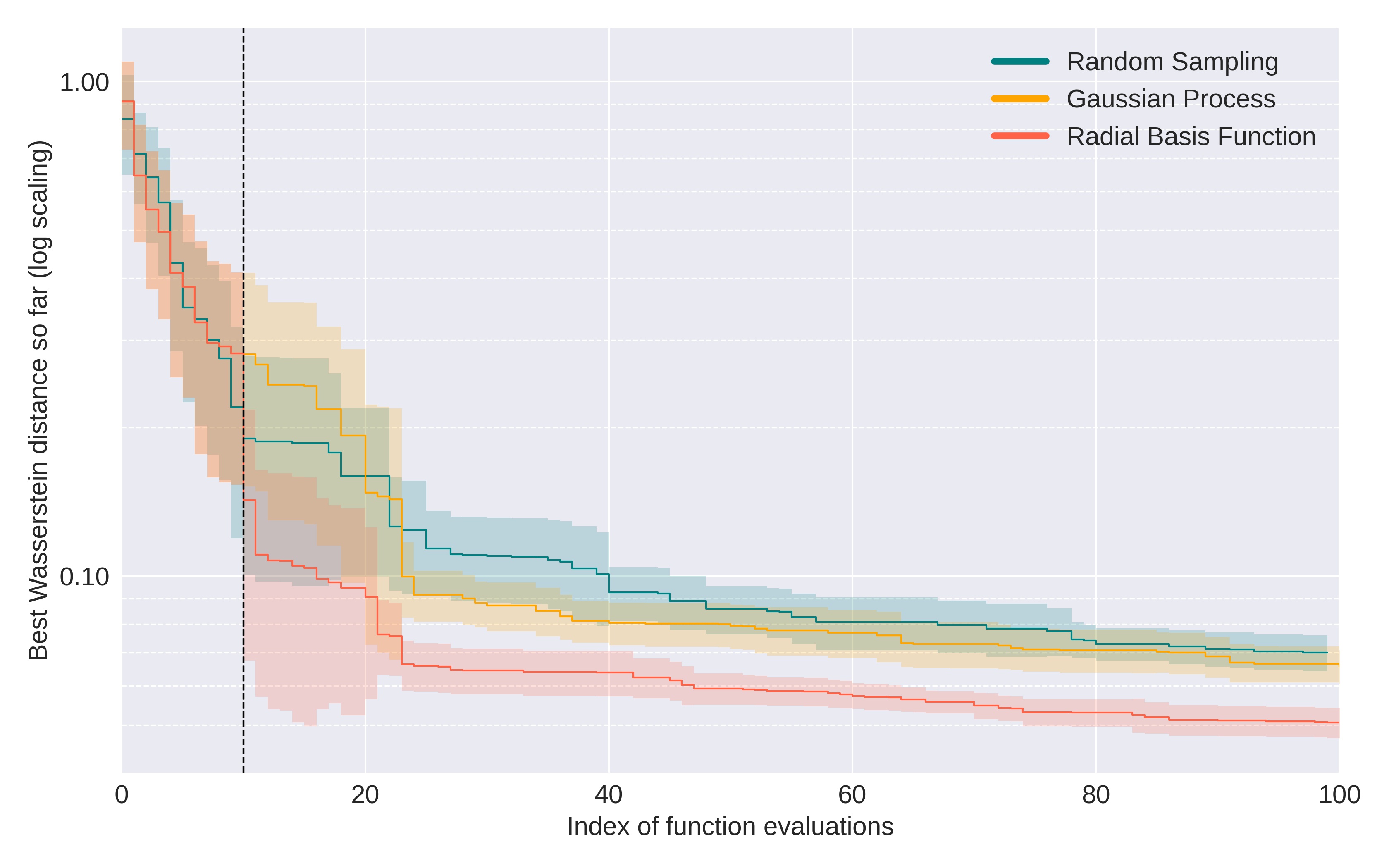}
\end{center}
\caption{Convergence plot showing the best Wasserstein distance (in log scale) found so far versus the number of hyperparameter evaluations done. Solid lines indicate mean value and shaded areas are standard deviations computed over the 20 independent HPO runs. Good solutions are found quicker  when using the RBF-based  optimization. The dashed black line represents the iteration at which the surrogate modeling starts, i.e., the end of the initial experimental design  (does not apply for random sampling).}
\label{fig:case1_convergence}
\end{figure}

The RBF-based HPO is also more efficient than the other two methods at finding good solutions as is evidenced in Figure~\ref{fig:case1_convergence}.   This figure shows  how quickly each approach finds improved solutions and converges.  To create this plot we recorded for each method and for each optimization run the best Wasserstein distance found so far. This results in a decreasing step function that we can plot against the function evaluation index (see data availability statement to access additional research materials and figures). For each method, we did 20 optimization runs, and thus we have 20 such step functions. We computed the mean and standard deviation across all 20 runs for each function evaluation index. In Figure~\ref{fig:case1_convergence}, we show the mean (solid line) and standard deviation (shaded area) for each method. Since we want to minimize the Wasserstein distance as fast as possible,  lower graphs indicate that better solutions are found quicker and we see that the RBF-based HPO outperforms both random sampling and the GP-based HPO. Note that the convergence graph for random sampling is only included for completeness and does not necessarily have much meaning since in random sampling the order at which samples are created and evaluated is somewhat arbitrary whereas the surrogate modeling based HPOs use the information of previously evaluated hyperparameter sets to make iterative sampling decisions. As indicated in Figure~\ref{fig:case1_solutions} (and also discussed in the next section), there are many solutions with similar performance. Thus, for this case study, random sampling has a relatively high chance at randomly selecting a decent  hyperparameter set.  However, we see that random sampling does not find as good solutions as the RBF-based HPO and, in particular, random sampling is not reliable at finding good solutions.  

\subsubsection{Sensitivity Analysis to Identify Important Hyperparameters}
\begin{figure}[ht]
\begin{center}
\includegraphics[width=\columnwidth]{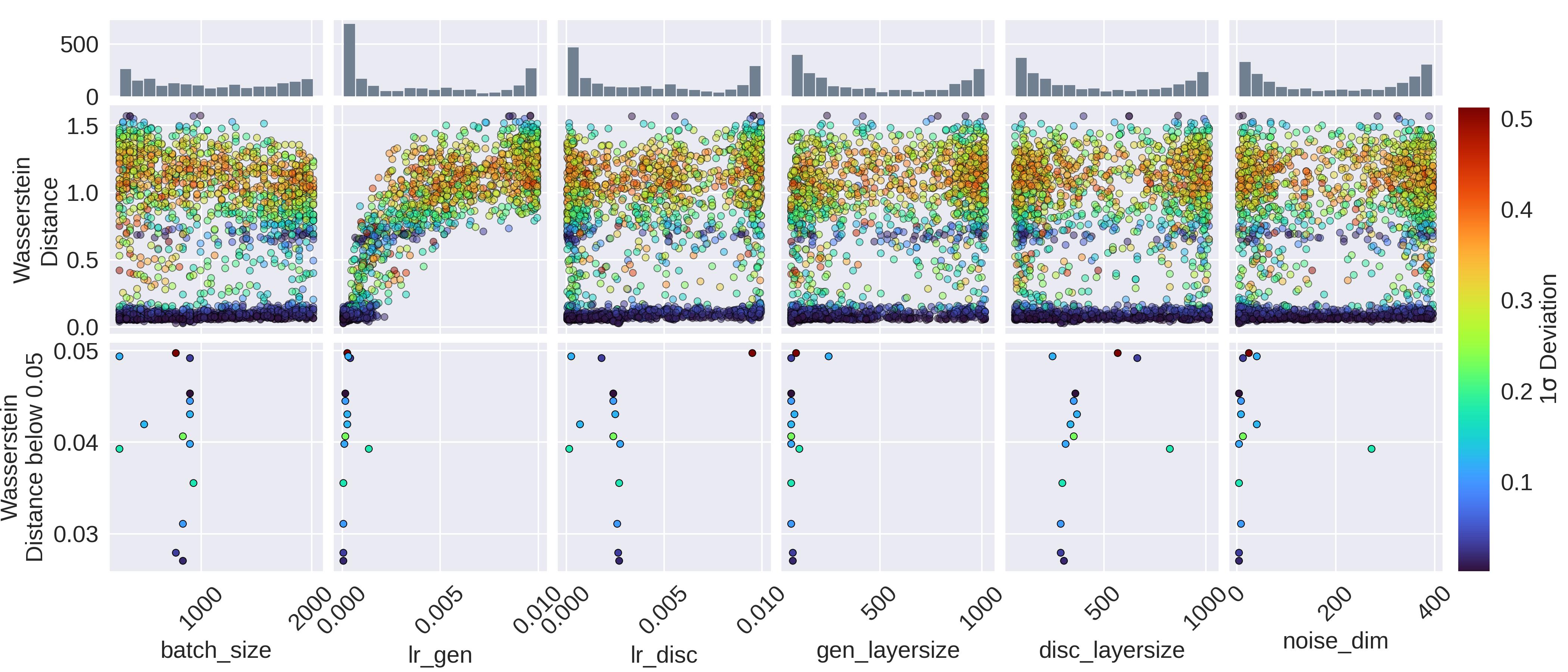}
\end{center}
\caption{Shown are the Wasserstein distances and $1\sigma$ deviations obtained  for all individual hyperparameters (second row) and the hyperparameter sets with a Wasserstein distance of less than 0.05 (bottom row).  The histograms in the  top row show where the samples were taken  in the hyperparameter space. Results are for RBF-based HPO. }
\label{fig:case1_sensitivity}
\end{figure}

In order to understand the importance of each  hyperparameter for the model performance, we display in Figure~\ref{fig:case1_sensitivity} for each hyperparameter the marginalized  2000 solutions from the RBF-based approach. Note that we are only considering the sensitivity of the model performance with respect to individual hyperparameters, i.e., hyperparameter interactions (second-degree sensitivities) and their combined impact are not shown.  The middle row of Figure~\ref{fig:case1_sensitivity} shows that the generator learning rate \texttt{lr\_gen} is correlated with the Wasserstein distance  as lower  generator learning rates yield  smaller  Wasserstein distances. Although potential second-degree correlations between hyperparameter settings are not represented, we can see that for the other five hyperparameters, the achievable solution quality can vary widely across. These results also indicate that there are potentially many different solutions with a very similar performance, i.e., as long as the generator learning rate is small, the values for the remaining hyperparameters may not matter as much. This would also explain the success of random sampling for finding decent solutions in almost every run.  

In the lower row of Figure~\ref{fig:case1_sensitivity}, we illustrate only those hyperparameters whose Wasserstein distance is less than 0.05 in order to investigate if good solutions cluster in subregions of the hyperparameter search space. We can see that for most hyperparameters, the good solutions are indeed found in a smaller subspace of the hyperparameter range, especially for the generator learning rate. It seems that GANs with a modest batch size, small generator learning rate, relatively larger discriminator learning rate, and a small noise dimension are generally favored. No significant improvement is seen when wider neural networks are used for the Generator and the Discriminator because smaller to moderate  layer sizes are favored. Based on these observations, we narrowed the search range in the focus search. 

The histograms in the top row of Figure~\ref{fig:case1_sensitivity} show how many solutions were sampled by the RBF approach across the range of  each hyperparameter. We can see that for  all hyperparameters, the sampling was more concentrated at the lower and upper bounds of the hyperparameter ranges. In particular, the generator learning rate was sampled predominantly at small values. The scatter plots in the bottom row of the figure indicate that for several hyperparameters, the best solutions were found toward the low end of the search space, which indicates that the HPO using RBF models is more likely to suggest hyperparameters that yield lower Wasserstein distances. 

\subsubsection{A Note on the Generator Learning Rate}

\begin{figure}[ht]
\includegraphics[width=\columnwidth]{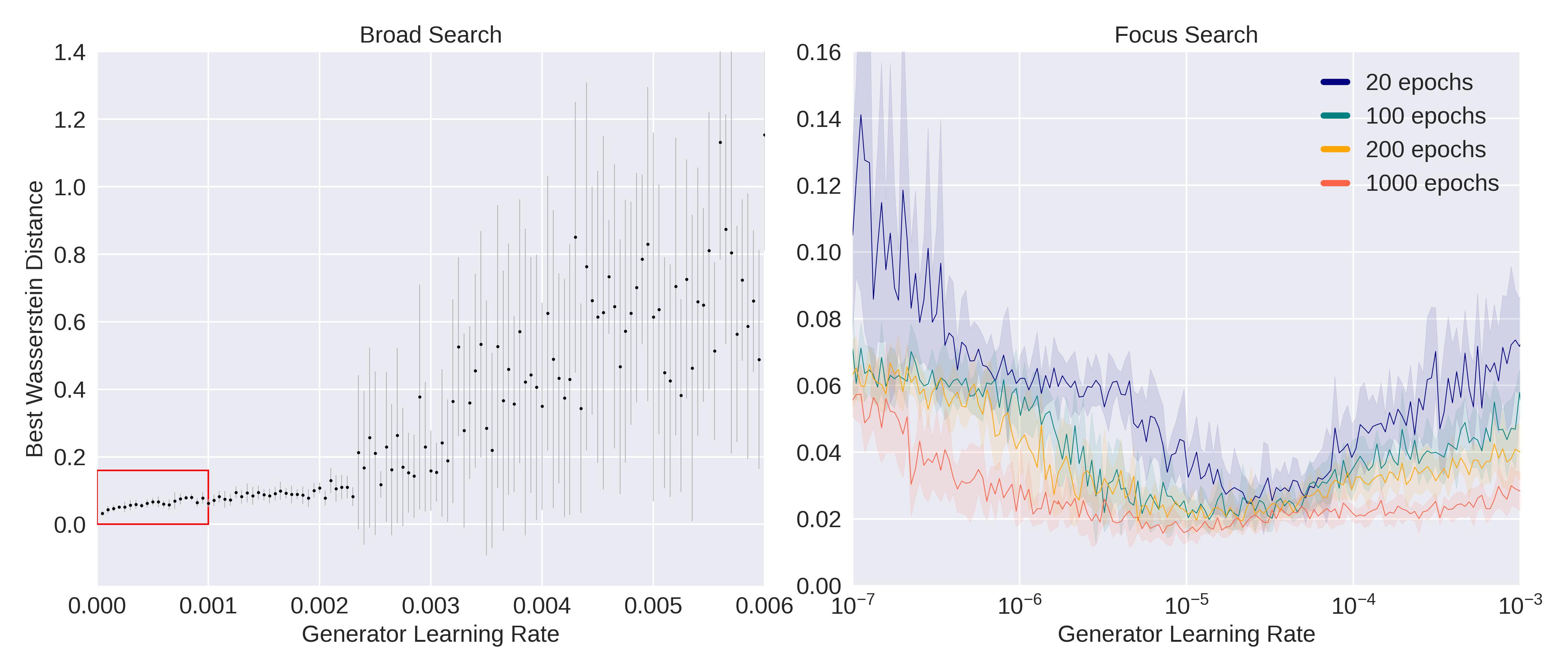}
\caption{Shown are  the best Wasserstein distances for different generator learning rates and their variability. The remaining hyperparameters were fixed to the values of the best model architecture found among the solutions shown in Fig.\,\ref{fig:case1_solutions}. The left plot shows the results from a broad search done in linear scale in (0, 0.006]. The right plot shows the results from a focus search done on logarithmic scale in $[10^{-7}, 10^{-3}]$.}
\label{fig:case1_lr}
\end{figure}

The results from the broad search (see Fig.~\ref{fig:case1_sensitivity}) revealed that the generator learning rate impacts the GAN's performance the most. Thus, we study in detail whether making the generator learning rate arbitrarily small necessarily leads to better model performance. To this end, we use the best-performing GAN architecture found during the RBF-guided optimization and we fix all hyperparameters except the generator learning rate, which we sample equidistantly in linear scale from the interval (0, 0.006]. We chose this range due to our observations shown in the bottom panel of the \lrgen \ column in Fig.~\ref{fig:case1_sensitivity}. A correlation study between the learning rate and the batch size was performed and no indications of a strong correlation were observed. In the left panel of Fig.~\ref{fig:case1_lr}, we plot the best Wasserstein distance and its  standard deviation against the generator learning rate. We can clearly see that the Wasserstein distance and its variability significantly increase for generator learning rates larger than 0.002, and thus it is interesting to study the performance for smaller learning rates (right panel).

In the right panel of Fig.~\ref{fig:case1_lr} we fix all hyperparameters and adjust only the generator learning rate, now sampling it equidistantly in logarithmic scale from $[10^{-7},10^{-3}]$. We  trained the corresponding models multiple times and illustrate the results (mean and variance over the trials). We show the results for different numbers of epochs for which the model is trained. Generally, we see that using more epochs yields better model performance  as is evidenced by the fact that the red graph (1000 epochs) is below all other graphs that use fewer epochs.

The advantage of using smaller learning rates and more epochs is that better solutions can be obtained, but on the other hand, large numbers of epochs lead to long model training times. Thus, trade-offs between model performance and required training time must be made. 

Figure~\ref{fig:case1_lr} shows that, for the considered application, we cannot make the generator learning rate arbitrarily small and expect improved model performance without increasing the number of epochs. There appears to be a performance optimum between [$10^{-6}$, $10^{-4}$], outside which the model performance deteriorates. We can also see that the vicinity of this optimum is narrower (in [$10^{-5}$, $10^{-4}$]) when using only few epochs (20, blue graphs) and it becomes wider as the number of epochs increases. Thus, for otherwise identical architectures, it is easier to find good model performance for a larger number of epochs since the optimum becomes less dependent on finding the exact learning rate. 

From the figure we can also see that the amount of improvement that can be achieved in the interesting  range of the learning rate decreases as the number of epochs increases. Thus, for a user of the GAN, a trade-off must be made between performance improvement and the required training time. If the GAN is to be trained only once on all available data, one can argue that large model training times are acceptable because solution quality is more important.  On the other hand, in settings where the GAN may need to be repeatedly retrained using  newly acquired  data, such as in  automated feedback loops, the GAN's training time may become the bottleneck and one may be content with somewhat less accurate models.  

\subsubsection{Focus Search and Final Results}

\begin{figure}[ht]
\begin{center}
\includegraphics[width=\columnwidth]{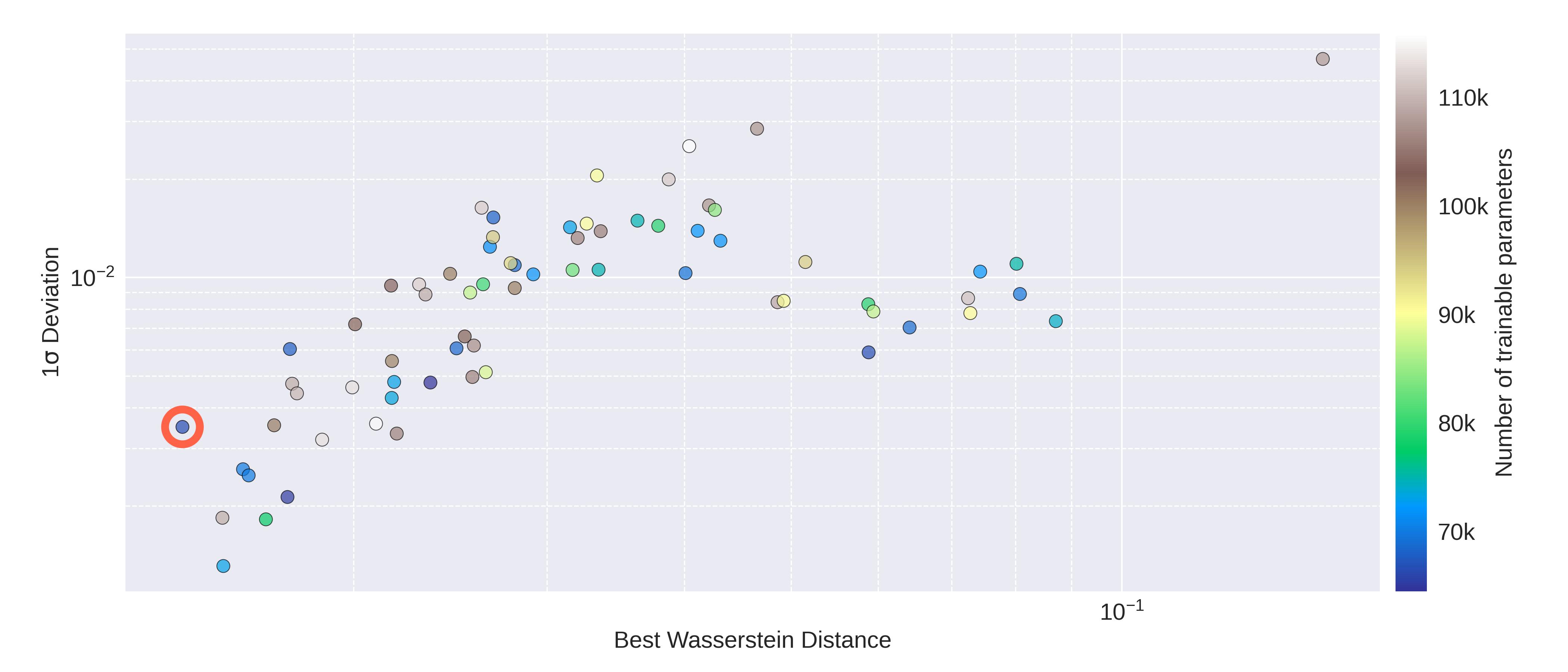}
\end{center}
\caption{All 70 solutions obtained with the focus search for case study 1 using the RBF based HPO. Shown are the Wasserstein distance and $1\sigma$ deviation for all hyperparameters that were sampled as well as the size (complexity) of the corresponding architectures in color. An optimal model was found with a best Wasserstein distance of 0.01398\,$\pm$\,0.00349 (circled in red).}
\label{fig:case1_focus}
\end{figure}

Based on our observation made during the sensitivity analysis after the broad search, we narrowed down the hyperparameter ranges to those listed in Table\,\ref{table:hps_range} under ``Study 1 - Focus search''. We repeated the HPO with the RBF surrogate model on this narrower range, training all the models over 1000 epochs. The results are presented in Fig.\,\ref{fig:case1_focus}. We show for each sampled hyperparameter set the corresponding Wasserstein distance and the $1\sigma$ deviation as well as the architecture complexity as illustrated in color. We can see that architectures of different complexities can reach different solution qualities. Depending on whether the goal is to select the solution with minimal Wasserstein distance, minimal $1\sigma$ deviation, or minimal complexity, different solutions would be chosen from this plot because no single point minimizes all three quantities at the same time. A multi-objective optimization formulation for the case when multiple quantities should be optimized simultaneously is outside of the scope of the current study.

For the studied hadronization processes application, we choose as best performing architecture the one with minimal Wasserstein distance encountered throughout our focus search (circled in red in Fig.\,\ref{fig:case1_focus}). This architecture has a generator and discriminator layer size of respectively 65 and 243 nodes, a generator and discriminator learning rate of respectively $4.0\times10^{-6}$ and $2.4\times10^{-3}$, a batch size of 642 events, and a noise dimension of 3. This architecture was trained eight independent times for 1000 epochs each and provides a best Wasserstein distance of 0.01398\,$\pm$\,0.00349. 

\begin{figure}[ht]
\begin{center}
\includegraphics[width=\columnwidth]{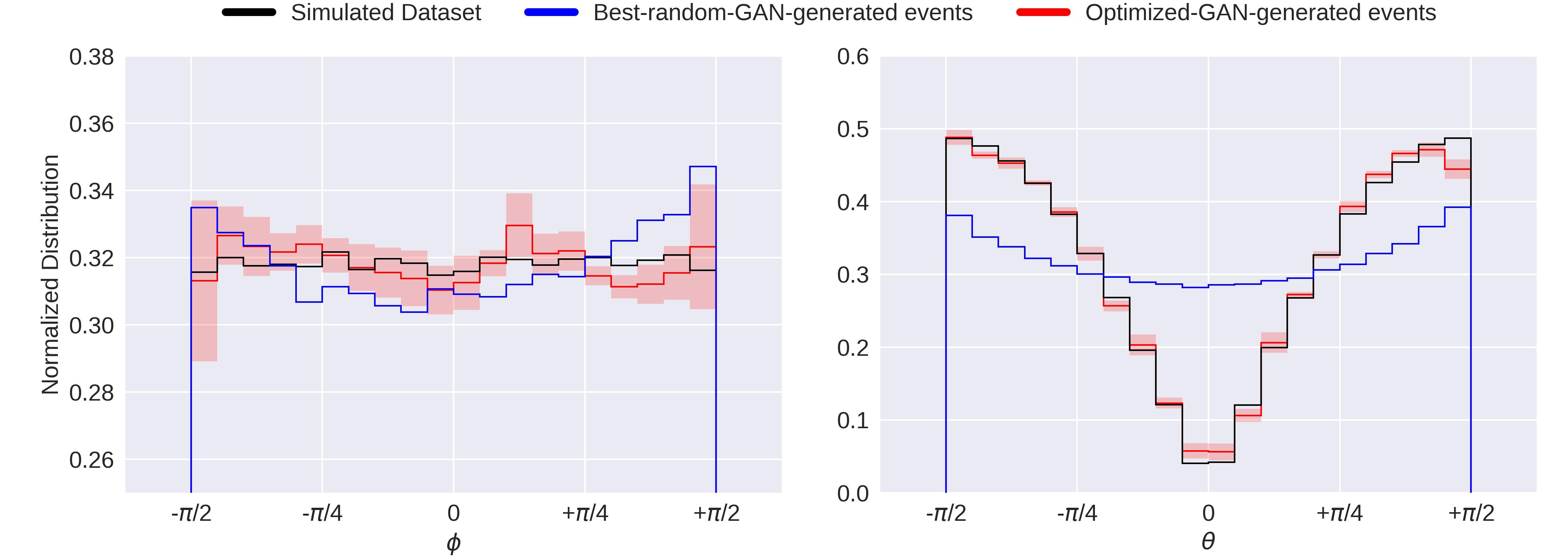}
\end{center}
\caption{GAN-generated distributions of the angles. The ground truth distributions are shown in black, the GAN-generated distributions from the best random model and the optimized GAN model are shown in blue and red, respectively. The shaded areas correspond to the variability of the predicted distributions.}
\label{fig:case1_histograms}
\end{figure}

We  ensemble the eight  predictions and we take  the mean values as the nominal predictions and the standard deviation as the uncertainty.  Fig.~\ref{fig:case1_histograms} shows the comparison among the simulated observables from \HW, the GAN-generated events from the best random GAN architecture found, and our optimized GAN-generated events.

\subsection{Case Study 2: Simulation of    \zmumu Events}
\label{dimuon}

Since the discovery of the Higgs boson at the LHC ten years ago, the search for the Higgs boson decaying into two muons has been the most promising analysis that can testify the interaction of the Higgs boson with the second generation fermions. The dominant Standard Model background of the search comes from the events where a $Z$ boson decays to a pair of muons, whose invariant mass distribution is modeled with parameterized functions. To estimate uncertainties of the background modeling, ATLAS uses the so-called spurious signal method~\citep{ATLAS:2012yve}, which requires massive amounts of simulated events so that uncertainties resulting from statistical fluctuations are negligible. In this study, we again use a GAN architecture to simulate the \zmumu events and we optimize the GAN hyperparameters with \hyppo.

Events from the \zmumu process are generated with \textsc{Sherpa} 2.2.1~\citep{Gleisberg:2008ta,Sherpa:2019gpd} at a center-of-mass-energy of 13 TeV with NNPDF3.0 NNLO set~\citep{NNPDF:2014otw}. 100000 dimuon events are generated and 5000 events are used for validation. The Generator is expected to learn the four momentum distributions of the two outgoing muons. In this case, no additional input is given to the Generator except for the noise values.

\subsubsection{Broad Search: Comparing Random Sampling and Surrogate Model Optimization}

\begin{figure*}[ht]
\includegraphics[width=\linewidth]{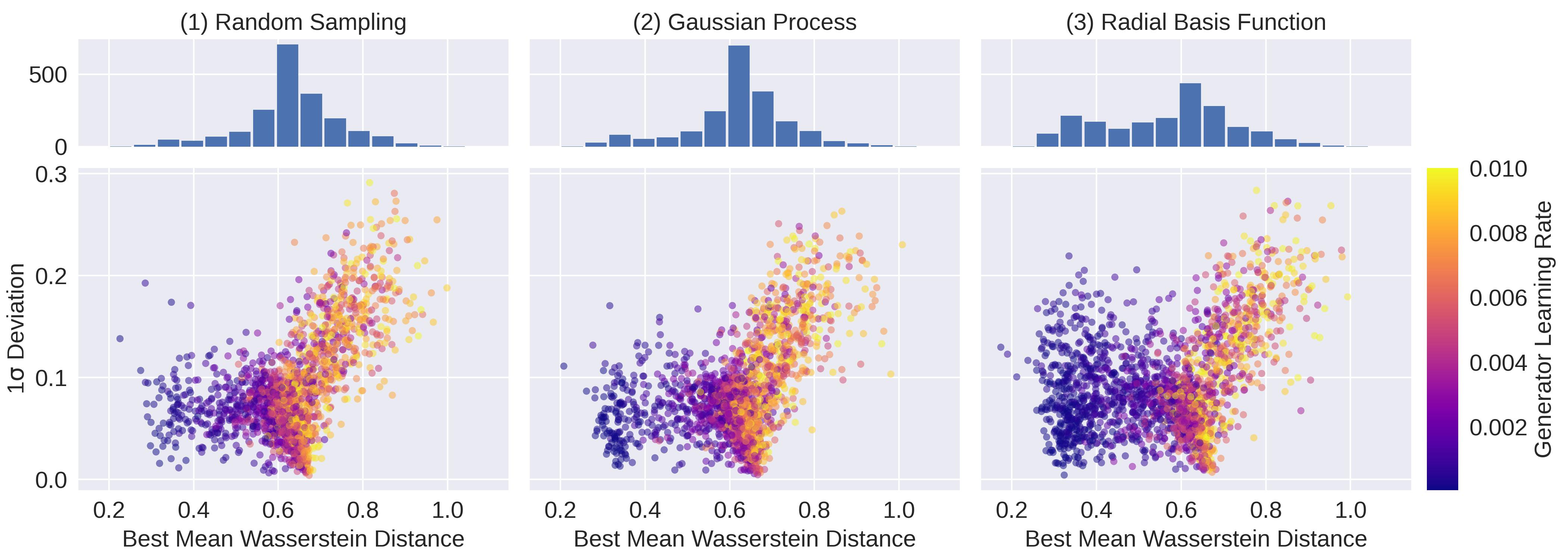}
\caption{HPO solutions for dimuon decay event simulations. Shown are the Wasserstein distances and $1\sigma$ deviations for all solutions sampled by each method. We prefer solutions in the lower left corner (low Wasserstein distance, low variability). The histograms indicate how many solutions were sampled by each method across Wasserstein distances. }
\label{fig:case2_solutions}
\end{figure*}

Similar to in the first case study, we performed 20 independent hyperparameter searches using each approach, i.e. random search and  both surrogate modeling methods (GP and RBF based, respectively). In Fig.\,\ref{fig:case2_solutions}, we show all solutions that were sampled with each approach and we observe that, similar to the hadronization case, the surrogate model based approach using radial basis functions  tends to provide more solutions with low Wasserstein distances as is indicated by the density of the point cloud toward the low Wasserstein distance regime and the histograms above the scatterplots.  The color scale  of the solutions  reflects the generator learning rate from which we can also observe that the smaller the generator learning rate,  the lower the Wasserstein distance of the model will be.

\subsubsection{Sensitivity Analysis to Identify Important Hyperparameters}

\begin{figure}[ht]
\begin{center}
\includegraphics[width=\columnwidth]{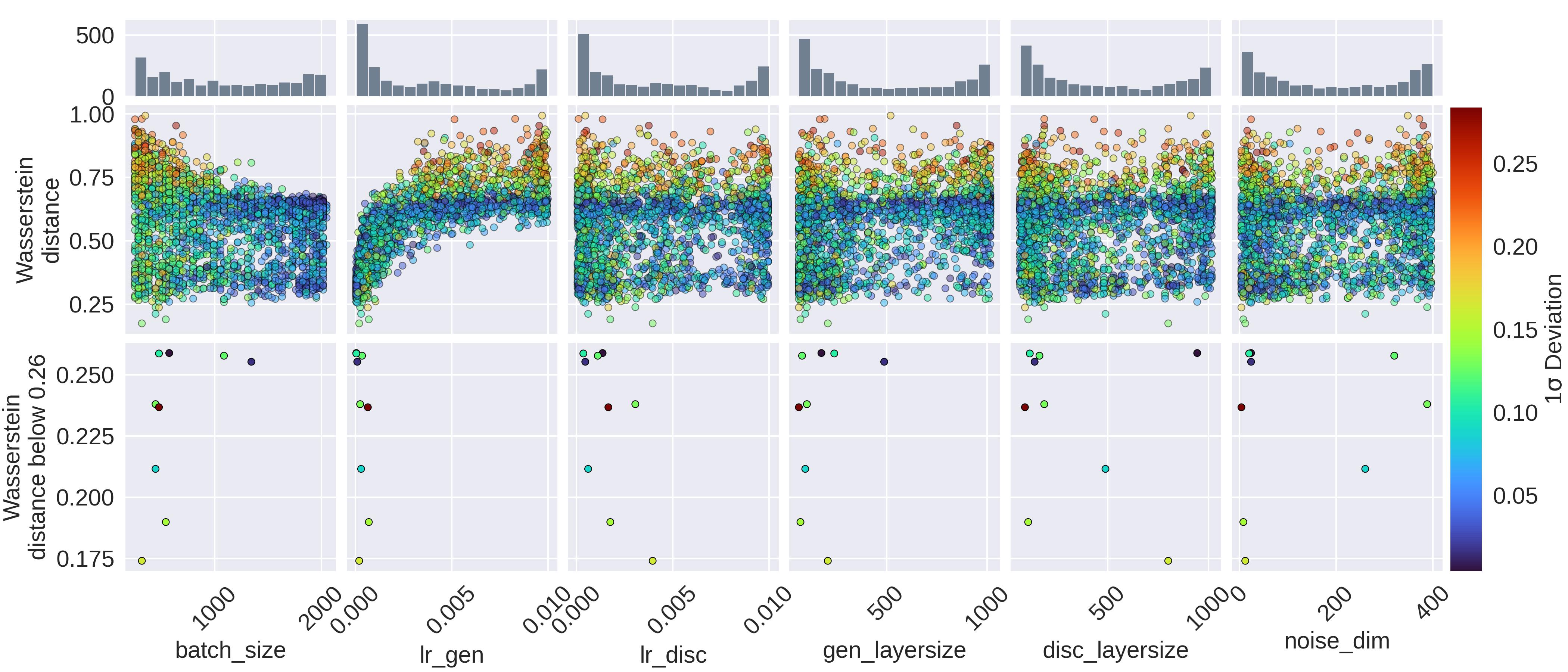}
\end{center}
\caption{Shown are the Wasserstein distances and $1\sigma$ deviations of the architectures sampled with the RBF-based HPO, marginalized for each hyperparameter. The bottom row shows the hyperparameter sets that achieved a Wasserstein distance of less than 0.26 and the histograms indicate where in the hyperparameter search space the samples were taken. }
\label{fig:case2_sensitivity}
\end{figure}

The results from the sensitivity analysis are shown in Fig.\,\ref{fig:case2_sensitivity} where we notice a  high-sensitivity pattern with respect to the generator learning rate similar to the first case study. For the remaining hyperparameters, we can  see that the range of Wasserstein distance values that can be achieved for a given hyperparameter value varies widely, and only for the batch size there appears to be a trend that larger values lead to generally lower worst-case Wasserstein distances. 
However, when plotting the hyperparameters that lead to  the best Wasserstein distances (less than 0.26, see bottom row images  of Fig.\,\ref{fig:case2_sensitivity}), we see that it is the smaller batch sizes that achieve the best performance. Similar to the first case study, also in this application  small generator learning rates are better. For the discriminator layer size and the noise dimension, the best solutions  cover a broader range of  values.  Thus, for the focus search, we could not reduce their ranges as much as for case study 1 (see Table~\ref{table:hps_range}, section Study 2 - Focus Search).  The histograms above the scatterplots are similar to case study 1, i.e., the RBF based HPO preferentially samples at the lower and upper bounds of the hyperparameter ranges. Given that many good solutions are found towards the lower and upper bounds, it appears that the RBF-based HPO method performs as expected. 

\subsubsection{Focus Search and Final Results}

\begin{figure}[b]
\begin{center}
\includegraphics[width=\columnwidth]{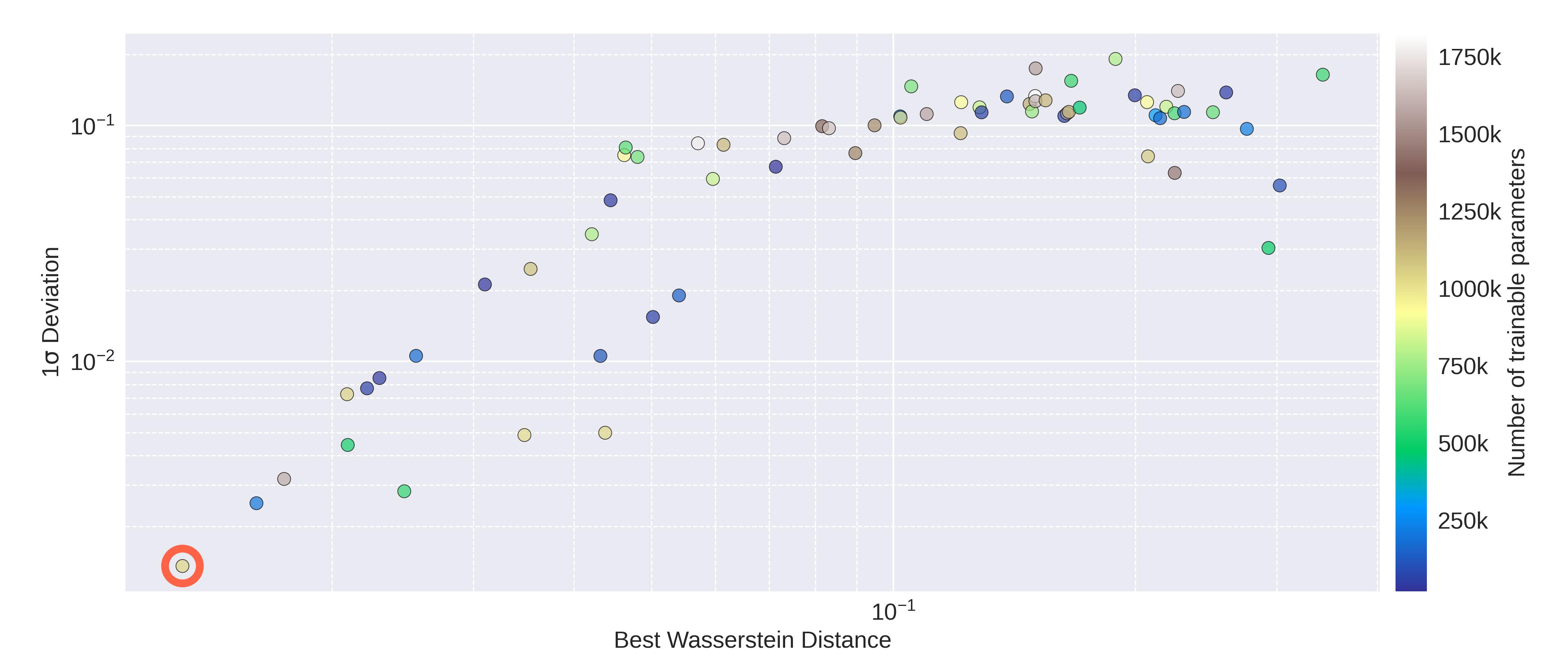}
\end{center}
\caption{All 67 solutions from the focus search with the RBF-based HPO. The best Wasserstein distance was found at 0.01304\,$\pm$\,0.00136 (circled in red), which is over an order of magnitude less than the best model found during the broad search.}
\label{fig:case2 _focus}
\end{figure}

Following the sensitivity analysis, we reduced the search space for the HPO  by targeting low generator learning rates and low data batch sizes. In Fig.\,\ref{fig:case2 _focus}, we present the final set of architectures sampled with the RBF-based HPO during the focus search. The best model (the solution in the bottom left corner of Fig.\,\ref{fig:case2 _focus}) achieved a  Wasserstein distance of 0.01304	with an associated standard deviation of 0.00136. The corresponding architecture has a noise dimension of 96, a batch size of 512, a generator learning rate of 9.5$\times10^{-5}$, a discriminator learning rate of 5$\times10^{-6}$, a generator hidden layer size of 960 nodes, and a discriminator hidden layer size of 104 nodes. Unlike in case study 1, there are no trade-offs to be made between the Wasserstein distance and the $1\sigma$ deviation because the best solution minimizes both metrics. If the model complexity was to be minimized as well, we would have to trade-off model performance and architecture size. 

\begin{figure}[ht]
\begin{center}
\includegraphics[width=\columnwidth]{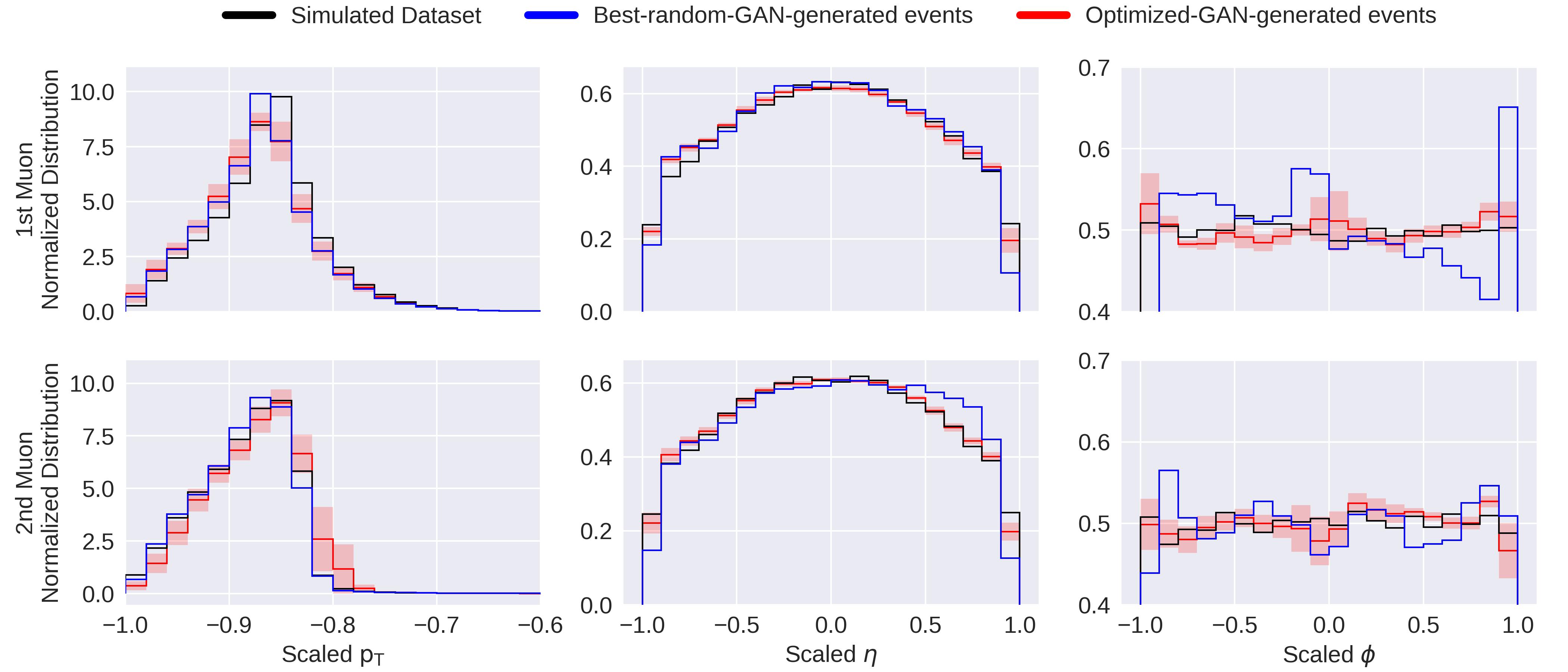}
\end{center}
\caption{Comparison of muon kinematics between the GAN-generated angles of both particles and the ground truth. The ground truth distributions are shown in black, the GAN-generated distributions from the best random model and the optimized GAN model are shown in blue and red, respectively. The shaded areas correspond to the variability of the predicted distributions.}
\label{fig:case2_histograms}
\end{figure}

In Figure~\ref{fig:case2_histograms}, we show the histograms of the ground truth and the GAN outcomes for the best architecture identified. We observe a good agreement between the GAN-generated distributions and the ground truth, indicating that a properly tuned GAN has great potential to be used as approximation model where  simulations that are based on parametrizations are computationally too expensive.

\subsection{A Note on Asynchronous Nested Parallelism}

Our hyperparameter optimization process relies on  multiple independent trainings of a large number of different hyperparameter sets. In order to speed up this process, we designed an Asynchronous Nested Parallelism (ANP) scheme \citep{hyppo_paper}. In this scheme, we evaluate the performance of different hyperparameter sets in parallel (first parallelization level), and we also parallelize the multiple trainings (trials) of each hyperparameter set (second / nested parallelization level). Both levels are executed concurrently using GNU parallel \citep{tange_2021_5523272}. Different hyperparameter sets correspond to different GAN model architectures, which may require significantly different training times due to different numbers of trainable parameters. In order to avoid a waste of compute time, we implemented asynchronism such that the surrogate model optimizer can start evaluating the next sample point immediately after any new architecture has its training completed thereby avoiding the wait for all the architectures trained in parallel to complete.

\begin{figure}[ht]
\begin{center}
\includegraphics[width=\columnwidth]{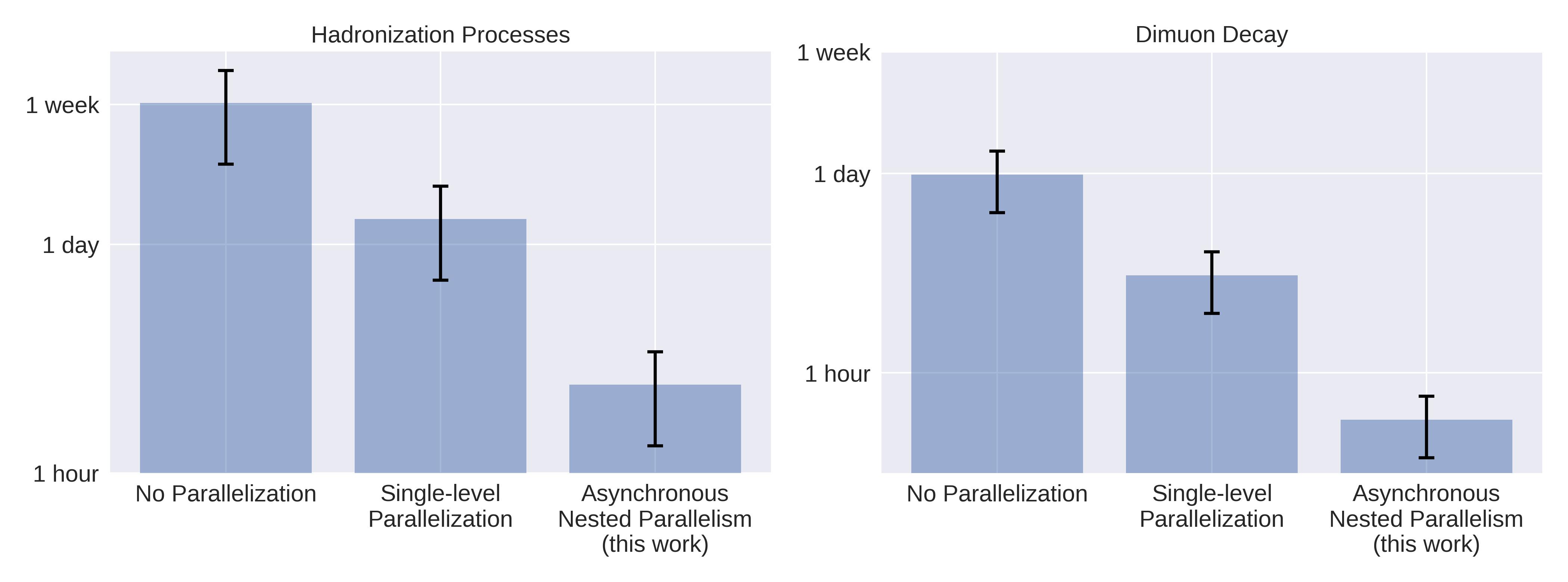}
\end{center}
\caption{Time requirement for HPO for both case studies. Shown are the time needed when parallelization is not employed, when one level of parallelization (execute multiple trials in parallel) is used, and when asynchronous nested parallelism is used. The processing times were extracted from the broad search with the RBF surrogate model over 100 architectures, with five trials per architecture, each  trained for 20 epochs. The  standard deviation is computed across all training times.}
\label{fig:speedup}
\end{figure}

Training one GAN architecture for 20 epochs as is done during the broad search can take anywhere between two minutes for a simple architecture with about 20000 trainable parameters, and $\sim$15 minutes for a complex architecture of about 240000 trainable parameters. However, if one wishes to train the architecture for a significantly larger number of epochs (say, 1000), it takes  several hours to complete. For example, an architecture of medium complexity with about 100000 trainable parameters will take about 3 hours. By leveraging HPC resources and our ANP scheme, one can therefore dramatically decrease the computing time required to complete an HPO run. Figure \ref{fig:speedup} shows for both case studies how long a single broad search HPO run would have taken if parallelization was not exploited, simple parallelization across repeated trainings (trials) was used, and asynchronous nested parallelism as described in \cite{hyppo_paper} was employed. In this work, we used up to 80 GPUs per HPO run, and we applied ANP to split the GPU resources into 10 sets of eight GPU processors where each set runs concurrently to execute multiple independent trainings of different architecture in parallel. Clearly, by using our ANP scheme, we were able to reduce the time required for HPO for the hadronization processes application from one week to three hours, and for the dimuon decay application from one day to 30 minutes.

\section{Conclusions and Future Research Directions}\label{sec:conclude}
We presented a detailed optimization and sensitivity analysis of the impact of the hyperparameters of GANs on the model performance in the context of simulating high energy physics events. For optimization, we employed a  variability-aware surrogate approach. We examined our  approach on two different physics problems: the simulation of hadronization processes and the simulation of the \zmumu process. We demonstrated a step-by-step workflow for how to best  tune  GAN architectures such that reliable prediction performance can be achieved. To this end, the HPO software \hyppo and its associated  sensitivity  analysis tools are used.

Given the intrinsic difference between the two physics problems, the GANs' hyperparameters may exhibit different sensitivities. However, we found that the generator learning rate appears to be the most important hyperparameter in both cases; a smaller generator learning rate often leads to better predictive performance. Therefore, more attention should be given to the generator learning rate during the tuning procedure.  In addition, the  surrogate optimization approach that uses radial basis functions  finds promising architectures more frequently and much better solutions than the random sampling approach in both cases. The choice of surrogate function is a key component for finding an optimal solution. We leave the exploration of other flexible and expressive surrogate functional forms for future work. 

There are multiple research directions that should be investigated in the future. From an HPO perspective, it would be interesting to address the architecture tuning problem from a multi-objective direction in which multiple performance measures (e.g., Wasserstein distance, variability, model complexity) are minimized simultaneously. In contrast to the presented approach, only trade-off solutions would be returned and potentially a larger variety of these solutions would be generated, allowing for a wider selection for the final solution. Our sensitivity studies are based on the first-degree analysis, i.e, the correlation between each hyperparameter and the performance metrics. It would be interesting for the future to take into account the correlations between hyperparameters in the sensitivity study. 
From a physics perspective, our case studies focus on simulations that generate a fixed number of observables. However, there are many  High Energy physics simulations that require the generation of a variable number of observables as is the case in  experimental data and developments in this space are needed. Also, a thorough sensitivity study similar to ours on other generative models such as variational autoencoders or normalizing flows that can produce a variable number of observables  would benefit the community significantly.

\section*{Author Contributions}
XJ generated the input data and implemented the GANs for the case studies and discussed the meaning of the results for the HEP applications.  VD conducted the numerical experiments using the GAN implementations  and also  implemented the analysis tools. JM guided the research and contributed to analyzing the results. All authors contributed to writing on all sections of the manuscript. They read and approved the submitted version.

\section*{Funding}
Vincent Dumont and Juliane Mueller are supported by the Laboratory Directed Research and Development Program of Lawrence Berkeley National Laboratory and the Office of Advanced Scientific Computing Research under U.S. Department of Energy Contract No. DE-AC02-05CH11231. Support for this work was also provided through the Scientific Discovery through Advanced Computing (SciDAC) program funded by U.S. Department of Energy, Office of Science, Advanced Scientific Computing Research and High Energy Physics for Xiangyang Ju.

\section*{Acknowledgments}
 This research used resources of the National Energy Research Scientific Computing Center (NERSC), a U.S. Department of Energy Office of Science User Facility located at Lawrence Berkeley National Laboratory, operated under Contract No. DE-AC02-05CH11231 using NERSC award ASCR-ERCAP0020826.

\section*{Data Availability Statement}

The entirety of the data\,\citep{dumont_vincent_2022_6975596} and codes\,\citep{ju_xiangyang_2022_6975552} used in this work have been made publicly available. Detailed instructions on how to access the data and reproduce the results presented here can be found in the online documentation accessible at \url{http://hpo-uq.gitlab.io/hyppo/papers/gan4hep}.

\bibliography{sn-bibliography}

\end{document}